\documentclass{desyproc}
\begin{document}
\title{Hidden Photons in Extra Dimensions}

\author{{\slshape Chris J. Wallace$^1$, Joerg Jaeckel$^{2}$, Sabyasachi Roy$^1$}\\[1ex]
$^1$Institute for Particle Physics Phenomenology, Durham University, United Kingdom\\
$^2$Institut f\"ur Theoretische Physik, Ruprecht-Karls-Universit\"at Heidelberg, Germany}

\contribID{wallace\_chris}

\desyproc{DESY-PROC-2013-XX}
\acronym{Patras 2013} 
\doi  

\maketitle

\begin{abstract}
Additional U(1) gauge symmetries and corresponding vector bosons, called hidden photons, interacting with the regular photon via kinetic mixing are well motivated in extensions of the Standard Model. Such extensions often exhibit extra spatial dimensions.
In this note we investigate the effects of hidden photons living in extra dimensions. 
In four dimensions such a hidden photon is only detectable if it has a mass or if there exists additional matter charged under it.
We note that in extra dimensions suitable masses for hidden photons are automatically present in form of the Kaluza-Klein tower.
\end{abstract}

\section{Motivation}

Extra U(1) gauge groups are a well motivated and consequently well studied extension of the Standard Model (SM). 
The simplest way the corresponding gauge bosons can interact with the Standard Model is via kinetic mixing with the photon~\cite{Holdom:1985ag}.
In absence of extra matter charged under this U(1) (and this is the case we are interested in here) this interaction is only observable if the hidden photon is massive.
Usually one can consider a Higgs mechanism\footnote{The extra Higgs may lead to additional constraints~\cite{Ahlers:2008qc}.}
or a Stueckelberg mechanism~\cite{Ruegg:2003ps} (and references of~\cite{Jaeckel:2010ni}).
If the hidden photon field extends into extra spatial dimensions an additional option presents itself in the form of a stack of massive Kaluza-Klein (KK) excitations (see also~\cite{McDonald:2010iq}).
In this note we will consider this option, and discuss its phenomenology and constraints which present some novel features compared to the canonical 4-dimensional hidden photon limits (for a review, see~\cite{Jaeckel:2010ni}).

\section{Toy Model}

We consider a simple toy model with a low energy effective theory defined by the action,
\begin{align}
S_{D} = \int \textrm{d}^D x \sqrt{g} \bigg( -\frac{1}{4}F^{\mu\nu}F_{\mu\nu}\delta^{d}(\vec{y}) -\frac{1}{4}X^{MN}X_{MN} \nonumber
-\frac{1}{2} \chi_{\rm D} F^{\mu\nu}X_{\mu\nu}\delta^{d}(\vec{y})\bigg),
\end{align}
which describes a $D$-dimensional bulk space (indices $M,N$)  in which gravity and a U(1)$'$ gauge symmetry live. The standard model and its U(1) is localized to a 3-brane (indices $\mu,\nu$) within the bulk at $\vec{y}=0$. 
Accordingly the kinetic mixing is also confined to the brane.
For the sake of low energy (w.r.t. the extra dimensional Planck scale) phenomenology, the brane possesses no inherent dynamics of its own, being infinitesimally thin and infinitely heavy. We take the extra dimensions to be flat. The theory and phenomenology of a hidden photon in a Randall-Sundrum setup in 5 dimensions has been treated in~\cite{McDonald:2010iq}.

Since no direct evidence for extra spatial dimensions exists, the $d=D-4$ additional dimensions have to be compactified.
We choose the most straightforward compactification onto a $d$-dimensional torus $\mathcal{T}_{d} = \mathcal{S}_1 \times \mathcal{S}_1 \dots \mathcal{S}_1$. Rough limits on the maximal size of the extra dimension are given in 
Table~\ref{tab:limits}.

As a consequence of the hidden photon field being constrained onto a torus the fields must possess appropriate periodicity. This allows us to decompose them in a Fourier series. For example in the simplest 5-dimensional case this reads,
\begin{equation}
X_{M}(x^{\mu},y^{a}) = \frac{1}{(\pi R)^{d/2}} \sum_{n>0} \left( X^{(n,+)}_{M}(x^{\mu})\cos{\left(\frac{n y}{R}\right)}
+ X^{(n,-)}_{M}(x^{\mu})\sin{\left(\frac{n y}{R}\right)}\right) +  \frac{X_{M}^{(0)}}{(2\pi R)^{d/2}}.
\end{equation}

Only the $X^{+}_{M}(x^{\mu})$ modes interact with the SM photon. The $X^{-}_{M}$ modes are associated with the sinusoidal part of the Fourier expansion and have zero amplitude for $y=0$, where we have localized the SM 3-brane. The expansion quickly becomes unwieldy for higher numbers of extra dimensions, but the result generalizes easily - only the terms in the expansion where all modes are $(+)$ (i.e. associated with the cosine) partake in kinetic mixing.

In the general case of $d$ extra dimensions, inserting this Fourier series and choosing an appropriate gauge we have 
\begin{align}
S_{\textrm{eff}} = \int d^4 x\,\bigg[ &-\frac{1}{4}F^{\mu\nu}F_{\mu\nu} -\frac{1}{4}X^{\mu\nu(0)}X_{\mu\nu}^{(0)} 
+ \sum_{n>0,p} \bigg(\frac{1}{4}X^{\mu\nu(n,p)}X_{\mu\nu}^{(n,p)} +  \frac{1}{2}\frac{n^2}{R^2}X_{\mu}^{(n,p)}X^{\mu\,(n,p)}\bigg)
\nonumber\\
&+ \sum_{n>0} \bigg(\frac{1}{2} \chi F^{\mu\nu}X_{\mu\nu}^{(n,+\ldots+)} \bigg)
+\ldots\bigg]
\label{eq:action}
\end{align}
\begin{wraptable}{r}{0.45\textwidth}
\centerline{\begin{tabular}{| c | c | c |}
\hline
$d$ & $1/R=m_0$ & $M^*$\\
\hline
1 & $>200$~$\mu$eV & $\gtrsim3\times10^5$~TeV \\
2 & $>700$~$\mu$eV & $\gtrsim3$~TeV \\
3 & $>100$~eV & $\gtrsim3$~TeV \\
4 & $>50$~keV & $\gtrsim3$~TeV \\
5 & $>2$~MeV & $\gtrsim3$~TeV \\
6 & $>20$~MeV & $\gtrsim3$~TeV \\
\hline
\end{tabular}}
\caption{Limits on the size of extra dimensions from precision tests of gravity. For $d=1$ the constraint arises from direct tests of the gravitational inverse square law~\cite{Hoyle:2004cw}. For $d=2-6$ the limits originate from constraints on the minimum value of the extra dimensional Planck scale $M^*$~\cite{Aad:2012cy}.}
\label{tab:limits}
\end{wraptable}
where the index $p$ is $d$-dimensional and denotes the combination of $+$ and $-$ modes from the Fourier expansion. The index $p$ is to be summed over, except in the case of the kinetic mixing term, which requires $p=+,\dots,+$ (all plus). Our initially $D$-dimensional field is separated into a stack of 4-dimensional hidden photon fields $X^{(n,p)}_{\mu}$. The dots indicate $d$ massless scalar fields originating from $X^{0}_{a=5\dots D}$, $d-1$ of which have associated stacks of KK modes (the missing stack being ``eaten'' by the now massive 4-dimensional hidden photon fields).
These scalars interact with the rest of the model only via gravity which we will neglect.
Note also that the 4-dimensional kinetic mixing parameter is suppressed by a volume factor compared to the higher dimensional mixing, $\chi = \chi_{\rm D}/(\pi R)^{d/2}$.

For our purposes the important consequence is that we now have a whole (infinite) tower of massive hidden photon fields with masses,
\begin{equation}
m^2_{\gamma'} = \frac{n^2}{R^2}(1+{\mathcal{O}}(\chi^2)) ,
\end{equation}
where $n$ is the KK mode number. Each of these interacts with the ordinary photon via the same kinetic mixing $\chi$. From the above we can see that we have observable massive hidden photons without the need to rely on an additional Higgs or Stueckelberg mechanism.

\section{Experimental and Observational Constraints}
For the simple 4-dimensional case a significant number of constraints already exists and have been discussed in the literature~\cite{Jaeckel:2010ni} (and references therein). Here, we will re-apply the same techniques to the case at hand. We will focus on three types of limits, stellar energy loss, fixed target experiments and precision measurements of $(g-2)$. A more detailed discussion with additional bounds will be presented in~\cite{inprep}.

\begin{figure}[!ht]
\centerline{\includegraphics[width=0.3\textwidth]{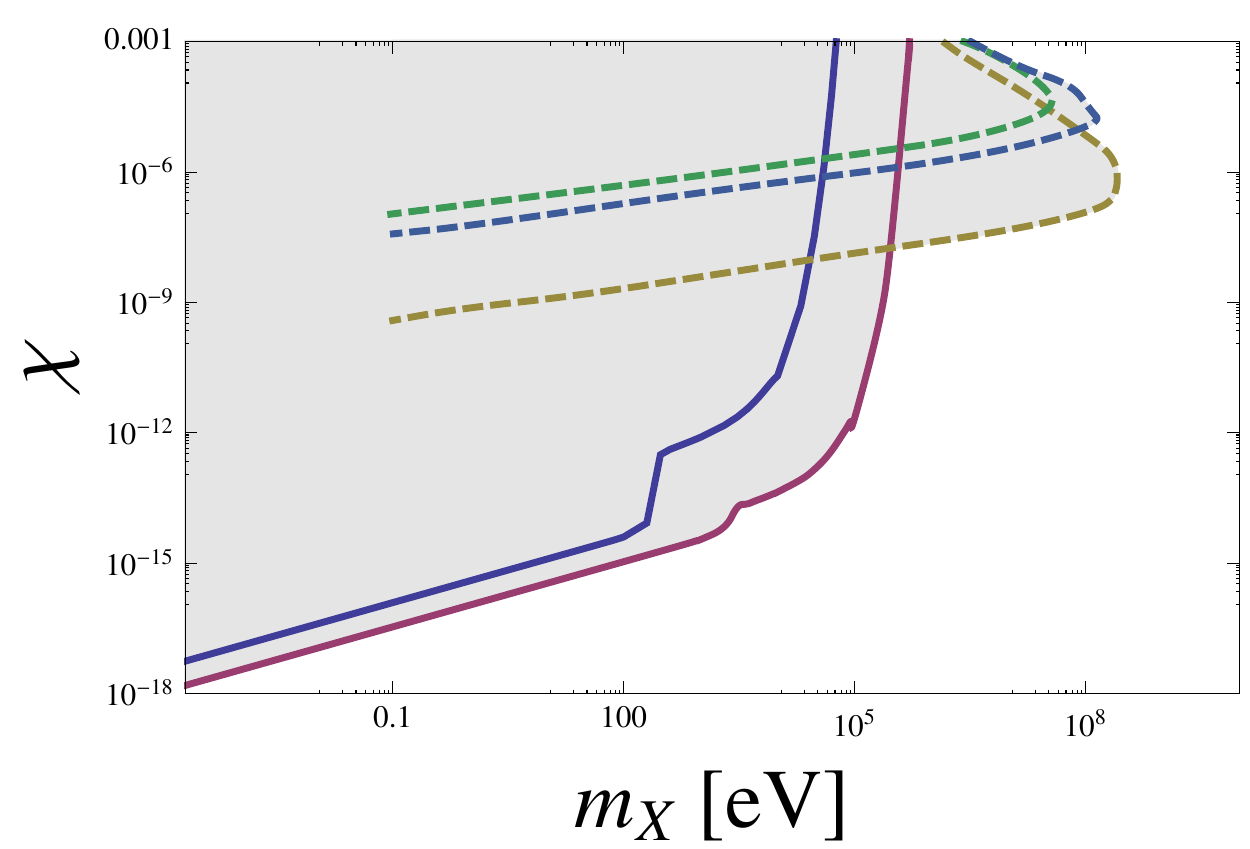} \includegraphics[width=0.3\textwidth]{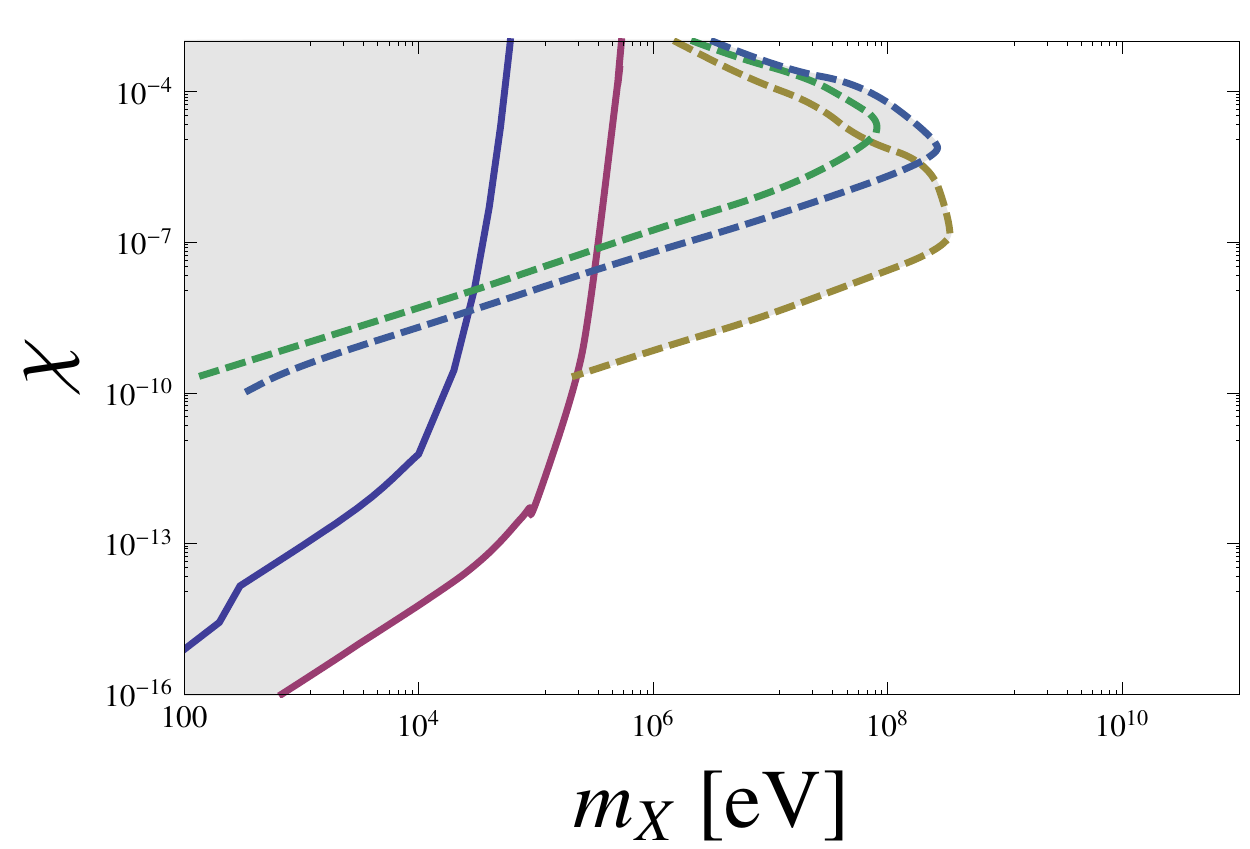}\includegraphics[width=0.3\textwidth]{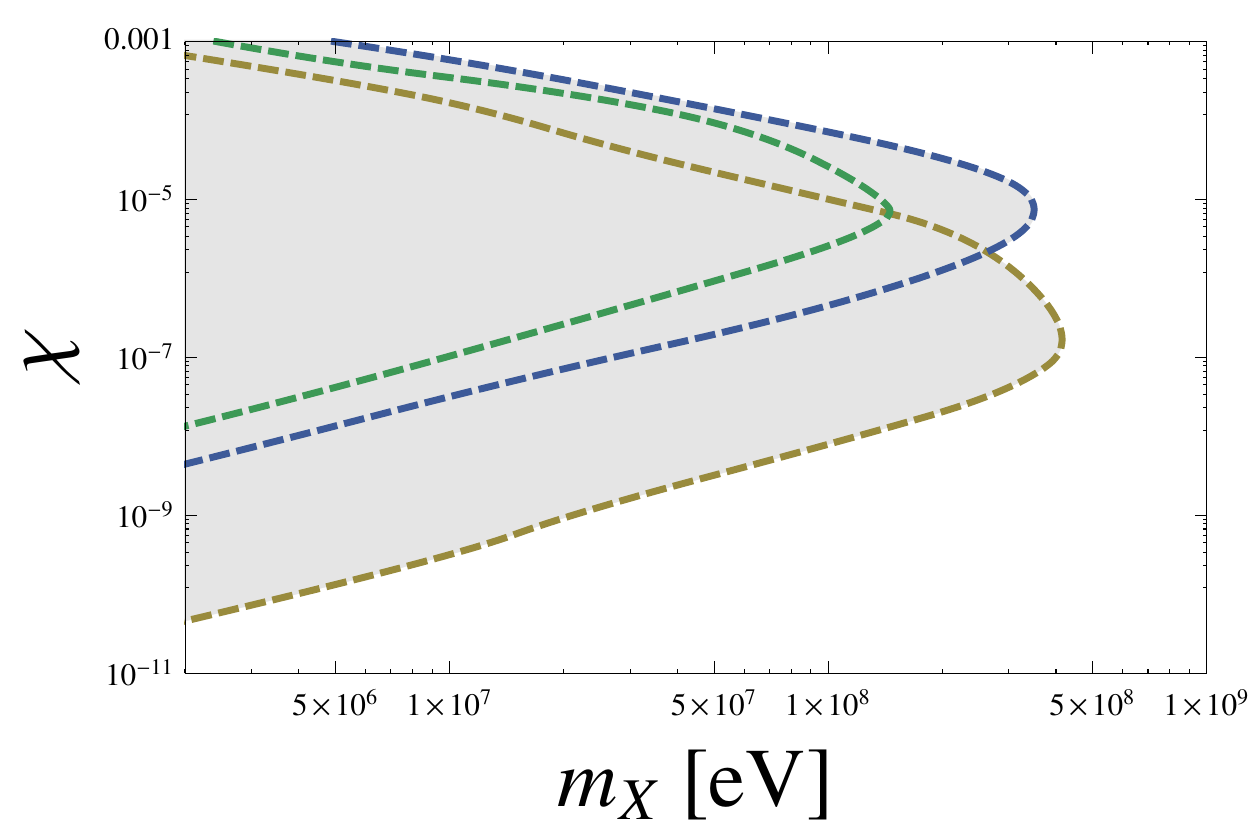}}
\caption{Limits from processes with real particle production (stellar energy loss and fixed target experiments), for $n=1,3,5$ extra dimensions.}\label{Fig:real}
\end{figure}

Stars lose energy when hidden photons are produced in the stellar interior and subsequently leave the star~\cite{Frieman:1987ui}.
If this energy loss is too great (typically more than the Standard Model luminosity), this is in conflict with observation.
In our extra-dimensional setup each KK mode constitutes a channel for energy loss, and the total energy loss is simply the sum over all channels. The resulting limits for $d=1,3,5$ are shown in Fig.~\ref{Fig:real} as greyed out areas, with solid blue and purple lines corresponding to solar and horizontal branch star energy losses, respectively.

\begin{wrapfigure}{r}{0.3\textwidth}
\centerline{\includegraphics[width=0.3\textwidth]{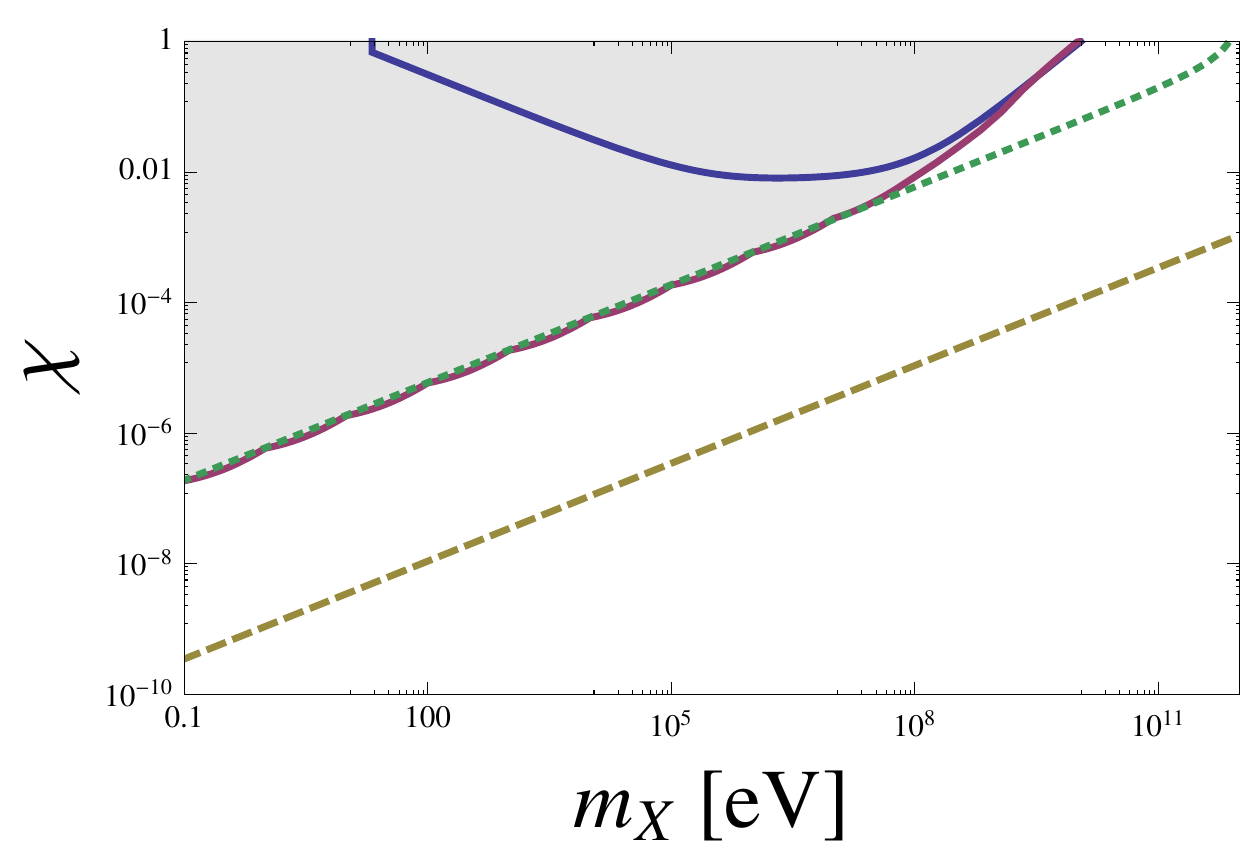}}
\caption{ $(g-2)$ perturbativity limits.}\label{Fig:virt}
\end{wrapfigure}

At higher masses fixed target experiments provide superior limits. 
Important experiments of this type are the E137, E141 and E774 beam dumps at SLAC and Fermilab~\cite{Bjorken:2009mm}. 
Essentially these experiments consist of an electron beam incident on a thick block of metal where hidden
photons are produced via Bremsstrahlung. Then follows a region of thick shielding which stops all Standard Model particles but not the hidden photons. Finally we have a volume where the hidden photons can decay into electrons and which is  instrumented to detect these electrons. Multiple hidden photon channels (KK modes) sum to give the total signal.
The corresponding bounds are shown as dashed lines in Fig.~\ref{Fig:real}.

Notice that the bounds have an interesting and generic feature that they become stronger as the mass of the lowest KK mode, $m_0$, decreases. In the general case, a small mass splitting gives a given experiment access to many modes to contribute to, say, energy loss. 

An interesting issue arises when computing the constraints on our toy model from processes where the hidden photon is produced off-shell. Taking the electron and muon $(g-2)$ constraint as an illustrative example, we soon encounter a problem with perturbativity.

The vertex correction responsible for the electron and muon anomalous magnetic moment is a 1-loop process. As such the 4-momentum in the loop is unconstrained and the whole, infinite, tower of KK modes of the hidden photon are accessible and contribute to the magnetic moment. 
Here we encounter the problem the non-renormalizabilty of higher dimensional gauge theories. To avoid this issue we impose a cutoff on the mass of the KK modes.

On top of that, a large number of KK modes contributing to the same quantity can invalidate our perturbative treatment.
To be on the safe side we require,
\begin{equation}
	\chi_{\rm pert.}^2 = \chi^2 \times \int_1^{\frac{\Lambda}{m_0}}d^d k
	=\chi^2 \times \int_1^{\frac{\Lambda}{m_0}}\frac{2\pi^{\frac{d}{2}}}{\Gamma(\frac{d}{2})} k^{d-1} dk
	\ll 1.
\end{equation}
Each choice of cutoff then defines a region where our perturbative treatment is valid, providing a severe limitation on the range of validity for the $(g-2)$ constraints.
This is illustrated for two different choices of the cutoff in Fig.~\ref{Fig:virt} - $\Lambda \sim M_*$ (dotted line) and $\Lambda \sim 1$~TeV (dashed).
The greyed out area would be excluded, but the whole grey region lies above the limit of a perturbativity for either choice of cutoff, and our perturbative treatment is insufficient.

\section{Conclusions}
Extensions of the Standard Model that contain extra hidden sector U(1) gauge bosons often also feature extra spatial dimensions. The hidden photon can then have its own KK tower and it can become observable even in absence of an additional mass generation via a Stueckelberg or Higgs mechanism.  We have presented exclusion limits on such a setup and we find that the limits on the allowed kinetic mixing are generically stronger than in the 4-dimensional case.\\
\\
{\bf Acknowledgements:} {\it We are grateful to the organizers of Patras 2013 and CJW is indebted to the ITP in Heidelberg for generous hospitality while working on this project.}


\begin{footnotesize}

\end{footnotesize}


\end{document}